\documentclass{astlb}
 
\usepackage{mathptmx}
\usepackage{txfonts}
\usepackage{pdflscape}
\usepackage[T1]{fontenc}
\usepackage{ae,aecompl}
\usepackage{graphicx}
\usepackage{changes}
\usepackage{hyperref}

\newcommand{\srg}{{\it SRG}}
\newcommand{\art}{ART-XC}
\newcommand{\ero}{{eROSITA}}

\newcommand{\sbat}{{{\it Swift}/BAT}}
\newcommand{\sxrt}{{{\it Swift}/XRT}}
\newcommand{\flx}{{ erg cm$^{2}$ s$^{-1}$}}

\begin{document}
\journalinfo{2022}{48}{7}{0}[0]
\title{Search for pre-burst emission from binary neutron star mergers with {\it Spectrum-Roentgen-Gamma}}

\author{ I.A. Mereminskiy\email{i.a.mereminskiy@gmail.com}\address{1}, A.A. Lutovinov\address{1}, K.A. Postnov\address{2,3}, V.A. Arefiev\address{1}, I.Yu. Lapshov\address{1}, S.V. Molkov\address{1}, S.Yu. Sazonov\address{1}, A.N. Semena\address{1}, A.Yu. Tkachenko\address{1}, A.E. Shtykovsky\address{1}, Z. Liu\address{4}, J. Wilms\address{5}, A. Rau\address{4}, T. Dauser\address{5}, I. Kreykenbohm\address{5} \\
{\it $^{1}$ Space Research Institute (IKI), Profsoyuznaya 84/32, Moscow 117997, Russia}\\
{\it $^{2}$ Sternberg Astronomical Institute, Moscow M.V. Lomonosov State University, Universitetskij pr., 13, 119992, Moscow, Russia}\\
{\it $^{3}$ Kazan Federal University, Kremlevskaya Str., 18,  Kazan, Russia}\\
{\it $^{4}$ Max Planck Institute for Extraterrestrial Physics, Gießenbachstraße 1, 85748 Garching, Germany}\\
{\it $^{5}$ Dr.\ Karl Remeis-Sternwarte \& Erlangen Centre for Astroparticle Physics, Sternwartstr.7, 96049 Bamberg, Germany}
}

\shortauthor{Mereminskiy et al.}
\shorttitle{Pre-merger X-ray emission of short GRB} 

\submitted{February 1, 2022}

\begin{abstract}
Close binary systems consisting of two neutron stars (BNS) emit gravitational waves, that allow them to merge on timescales shorter than Hubble time. It is widely believed, that NS-NS mergers in such systems power short gamma-ray bursts (GRB). Several mechanisms which could lead to electromagnetic energy release prior to a merger have been proposed. We estimate the ability to observe the possible pre-burst emission with telescopes of {\it Spectrum-Roentgen-Gamma}. We also investigate first such event, GRB210919A, which  fell into the field of view of the SRG telescopes less than two days before the burst.

{\sl Keywords:\/} surveys, X-ray sources, gamma-ray burst
\end{abstract}


\section{Introduction}

Short gamma-ray bursts (SGRBs) are narrow pulses of X-ray and gamma-rays, lasting typically for less than a second (up to several tens of seconds in some extreme cases, \citep{rastinejad22})  which constitute a significant, although lesser part of the total gamma-ray burst population \citep[see, e.g.][]{mazets81,kouv93,svinkin16,gbm4}.

Thanks to the recent simultaneous gravitational wave (GW)/gamma-ray detection of such an event GW170817 \citep{abbot17_GW} (see, also, \citealt{pozanenko20} for the claimed detection of the second similar event S190425z) the origin of at least some part of SGRBs are now secured. They are produced during neutron star - neutron star (NS-NS) mergers in binary systems: rapid inspiral generates characteristic ''chirp'', observed in GWs, while relativistic jets from a newly-born black hole (BH) powers the observed $\gamma$- and X-ray emission \citep{rezzolla11,ruiz16}. At later stages the optical transient (so-called "kilonova" \citealt{li98,metzger19kn}) are expected to arise, powered by a radioactive decay of neutron-rich nuclei, which condensed from the NS debris matter. It should be noted, that there was proposed another possible mechanism that explains electromagnetic emission from closing in NSs. In ''tidal stripping'' scenario \citep{clark77,blinnikov84} instead of merging,  part of the matter from  one of the NS is accreted onto the other. As a result of this accretion, one of the NS loses matter until its mass reaches critical value and then it explodes \citep{blinnikov21}. 

One of interesting possibilities in tight NS-NS binaries (BNS) is that there could be mechanisms that lead to the energy release prior to a merging. In fact, at least on short timescales ($\approx 1$ s before GRB) such events -- so-called precursors -- are observed \citep{koshut95}, although rarely, less then for 1\% of SGRBs \citep{minaev17}.  The origin of these precursors are not clear, although some authors connect them to crustal failures, caused by mutual tidal interactions \citep{tsang12, suvorov20}, while others suspect interactions between NS magnetospheres \citep{hansen01,lai12,metzger16,wang18}. In case of ''tidal stripping'' precursor emission could also be caused by increase of accretion rate onto a more massive NS before its low-mass counterpart loses its stability \citep{blinnikov21}. 

There is also a proposed class of common envelope jet supernovae \citep[CEJSN; ][]{gilkis19} in which BNS could merge inside the envelope of a red giant star (RGS). This type of transients could produce a bright emission in broad wavelength range from optical to X-rays for months before the merging \citep{soker21}. Although, it is necessary to note that in order to produce a ''classical'' short $\gamma$-ray burst in this scenario some fine-tuning is required for the merging to occur close to the RGS surface. 

Due to its unpredictable occurrence, no strong upper limits on the pre-merger emission from BNS on longer timescales (days to thousands of seconds) have ever been published. In this Letter we show that the \srg \citep{srg} observatory have a sufficient chance to observe BNS in their last hours-days before the merging. We also discuss short GRB210919A, which was observed by \srg\, less than two days before the burst.

\begin{figure*}
\centering
   \includegraphics[width=0.6\textwidth]{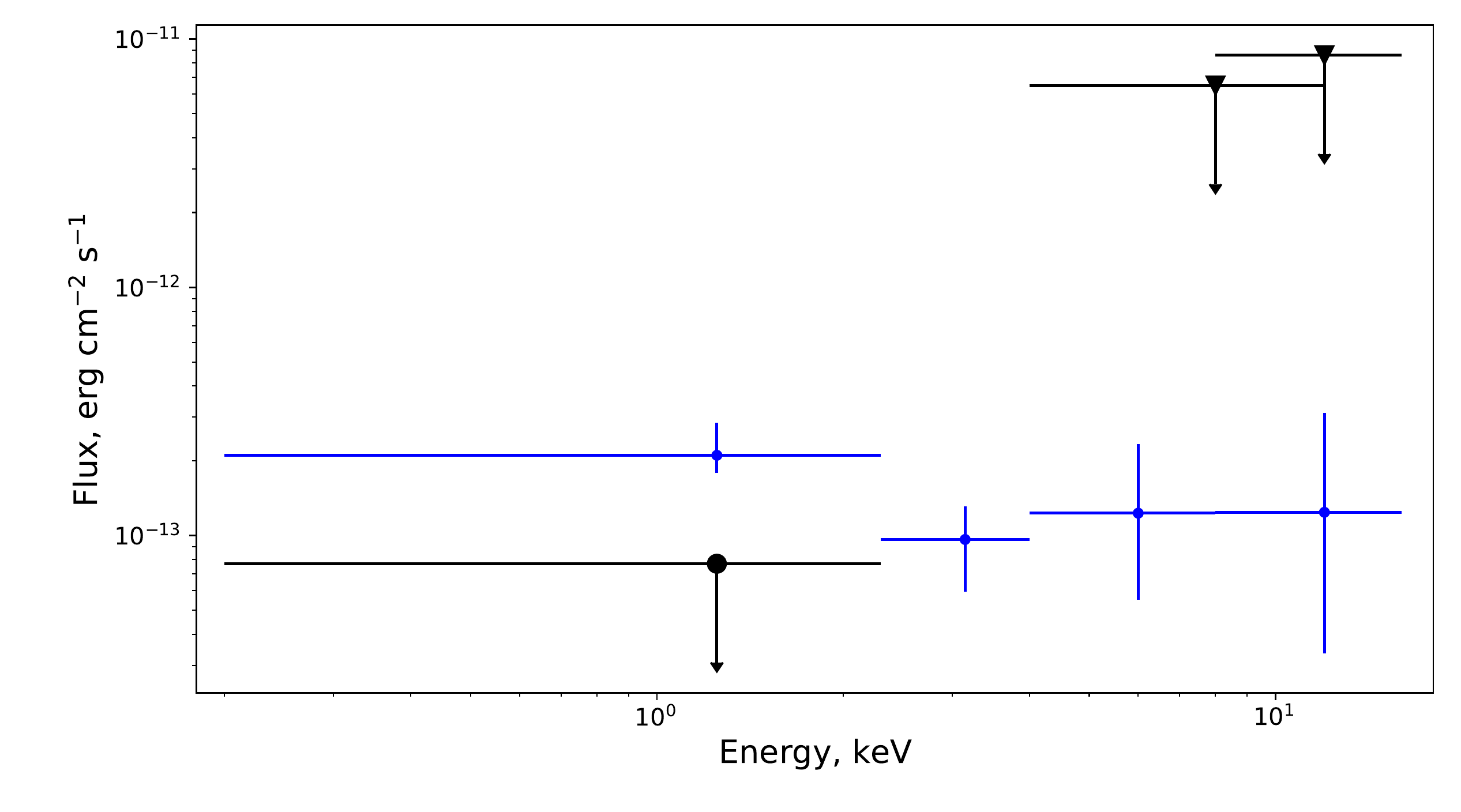}
     \caption{ Energy dependent upper limits on the pre-merger emission of GRB210919A obtained with \srg\, telescopes: \ero\, data shown with circles, \art\, with triangles. Model flux from best-fit of {\it Swift}/XRT early afterglow spectrum shown with blue points.}
     \label{fig:spe}
\end{figure*}

\section{Possibility to observe pre-merger emission from BNS with {\it SRG}}

Observations of BNS in last days before the merging can be only serendipitous. Given its survey strategy, \srg\, is the most suitable observatory for such observations. Covering about 1\% of sky each day, the {\it Mikhail Pavlinsky} ART-XC \citep{pavlinsky21} and eROSITA \citep{predehl21} telescopes have the highest chance to observe the position of upcoming transient among the sensitive grazing-mirror telescopes.

However, in order to recognise the observed X-ray transient as a pre-merging event, some kind of an external trigger is needed. It could be the detection of a well localised SGRB or a kilonova in the optical/near-IR band. 
The current generation of space-based gamma-ray burst monitors detects $\approx 40$ SGRBs per year, although only {\it Swift} can provide accurate enough positions to search for the  corresponding X-ray transients. 
The latest \sbat\, GRB catalog\footnote{\url{https://swift.gsfc.nasa.gov/archive/grb_table/index.php}}, \citep[see, ][for details]{sgrb3} suggests that over nearly 17 years of observations approximately 90 SGRBs were localised with \sxrt, given an estimated rate of 5.3 events per year. 
Optical surveys, such as {\it Zwicky Transient Facility} \citep[ZTF, ][]{ztf} or upcoming Large Synoptic Survey Telescope (LSST, \citealt{lsst}) could provide additional targets, including off-axis events (such as GW170817 \citet{margutti17}), that are biased against the detection by $\gamma$-ray monitors, due to their faintness. \citealt{andreoni21} predicts that LSST could find 0.3-3.2 kilonovae per year with a specifically tailored observational program. No viable candidates were found in 23 months of ZTF data \citep{andreoni20}. 

Therefore, it is straightforward to estimate that during its four-year survey, \srg\, will serendipitously cover the position of $\approx$0.2 upcoming SGRBs inside one-day window before the merging. This rough estimate agrees well with earlier estimates on the sGRB afterglow detection rate of $\approx 0.1$ yr$^{-1}$ \citep{khabibullin12grb}.

Moreover, we could estimate total number of such transients detected in all-sky survey in local Universe (neglecting cosmological effects) by fixing pre-burst luminosity at one day before the burst and also assuming that emission is isotropic. If X-ray luminosity of the pre-burst emission is $L_\mathrm{X,42}$ (in units of 10$^{42}$ erg s$^{-1}$) in 0.2-2.3 keV band, in which eRosita have typical sensitivity of $F_\mathrm{X} = 10^{-13}$ \flx in a day, $A_\mathrm{sky}=360/41253\approx 0.009$ -- part of sky, covered per day and  $R_\mathrm{NS-NS} = 10-1700$ Gpc$^{-3}$ year$^{-1}$  is a BNS merger rate (measured during first three observing runs with  LIGO-Virgo \citep{ligo21}). Then, in volume of $V=0.1 L_\mathrm{X,42}^\frac{3}{2}$ Gpc$^3$ we could expect to see $N_\mathrm{observed} \approx V\times A_\mathrm{sky} \times R_\mathrm{NS-NS}$ events year$^{-1}$. For pre-merger luminosity of 10$^{42}$ erg s$^{-1}$ we, therefore, could expect between $0.01..2$ events per year. 

However, it should be noted, that it would be tricky to distinguish such events from other transients, routinely seen by {\it SRG}, such as flares on nearby stars, AGN variability, e.t.c.

\section{GRB210919A}

The GRB210919A was first detected by \sbat\, on 00:28:33 UT, September 19, 2021 \citep{2021GCN.30846....1T}, and was soon observed with \sxrt. This observation allowed to obtain the precise localisation of the soft X-ray afterglow with coordinates of RA, Dec = 80.25448, +1.31153 (FK5, J2000, the 90\% confidence radius is 4.6\arcsec, \citealt{2021GCN.30850....1G}). Follow-up observations in optical/near-IR wavebands found no bright transient sources inside \sxrt\, error region \citep[][e.t.c.]{2021GCN.30852....1P, 2021GCN.30858....1Z, 2021GCN.30860....1G, 2021GCN.30868....1P, 2021GCN.30883....1K, 2021GCN.30934....1O, 2021GCN.30983....1K}, however, a single weak NIR source was observed (with $i'= 24.14\pm0.30$, $R_{c} = 24.47\pm0.53$ magnitude in AB system \cite{2021GCN.30883....1K,2021GCN.30884....1K}), which soon faded \citep{2021GCN.30983....1K}. Deep imaging of the GRB field revealed two galaxies with 20.5 and 24 magnitude in  $r$-band \citep[AB, ][]{2021GCN.30934....1O}, located at same redshift $z=0.2411$ \citep{LBTz22}. Projected distance between the faded optical source and these galaxies is 13 or 50 kpc, assuming that the optical transient lies on same redshift. Such distances from host galaxy are typical for short GRBs \citep[see, e.g.][ and references therein]{fong13,berger14}, and usually explained by velocity kicks received during the supernova explosion of one of the components \citep{fryer97}. All of this lead us to proposition, that the observed NIR transient was an afterglow of GRB210919A and the merger happened in group of galaxies at $z=0.2411$. A deep {\it Chandra} observation performed $\approx 2.2$ days after the burst also failed to detect an afterglow \citep{2021GCN.30879....1S}.

In order to show the temporal evolution of the GRB X-ray emission, we translated all observed fluxes to the 0.3-10 keV energy band, assuming that at every moment the event spectrum is described by a simple absorbed power-law model with an absorption column thickness of $1.6\times10^{21}$ cm$^{-2}$ \citep{galnh13}. For the main impulse we used $\Gamma=1.58$\citep{2021GCN.30863....1B}, as it was measured by \sbat. \sxrt\, data were processed with the online analysis tool \citep{evans09}. During the prompt \sxrt\, observation the afterglow was detected with the flux of $4.5\times 10^{-13}$ \flx. The source spectrum was consistent with the absorbed power-law with the absorbing column thickness close to the Galactic value and a power-law index of $2.1^{+1.4}_{-1.2}$. 
There were several late-time follow-up observations: two by \sxrt, started approximately $5$ ks and $280$ ks after the burst, and a rapid {\it Chandra} Target-of-opportunity observation, that lasted for 20 ks and started $180$ ks after the burst. In all of these observations afterglow was not detected, with the stringent $3\sigma$-upper limit on the source flux of $7.5\times^{-15}$ \flx\, in 0.3-10 keV band \citep{2021GCN.30879....1S}, assuming that the spectrum has not changed after the first afterglow detection.

The \art\, telescope covered the sky field of GRB210919A two days prior to the burst, with a mean time of about 1.9 days before the burst. We produced the calibrated event lists and sky images using the {\sc artproducts} pipeline v0.9 with the {\sc caldb} version 20200401. No source was detected in the standard 4-12 keV energy band, nor in the harder 8-16 keV one. The corresponding 95\% upper limits on the source flux are $6.5\times10^{-12}$\flx\, and $8.6\times10^{-12}$\flx , respectively, assuming a Crab-like spectrum. 

The field of GRB210919A was visited 8 times by \ero\, with a total exposure time of $\sim261$\,s during eRASS4, starting from 13:13:46UTC on September 16, 2021. Given its larger field of view, the last \ero\, visit of the source occurred significantly later than for ART-XC, on 17:13:48UTC, September 17, 2021. We processed the data with the standard \ero\ Science Analysis Software System (eSASS, version eSASS\_users201009) \citep{brunner18esas} pipeline. Assuming that the spectrum is an absorbed power-law with a column density of $1.6\times10^{21}~\mathrm{cm}^{-2}$ and a photon index of $\Gamma=1.9$, we obtained a $3\sigma$ upper limit of $7.7\times10^{-14}$ \flx\, in the soft 0.2-2.3 keV band. 

\begin{figure*}
\centering
   \includegraphics[width=0.6\textwidth]{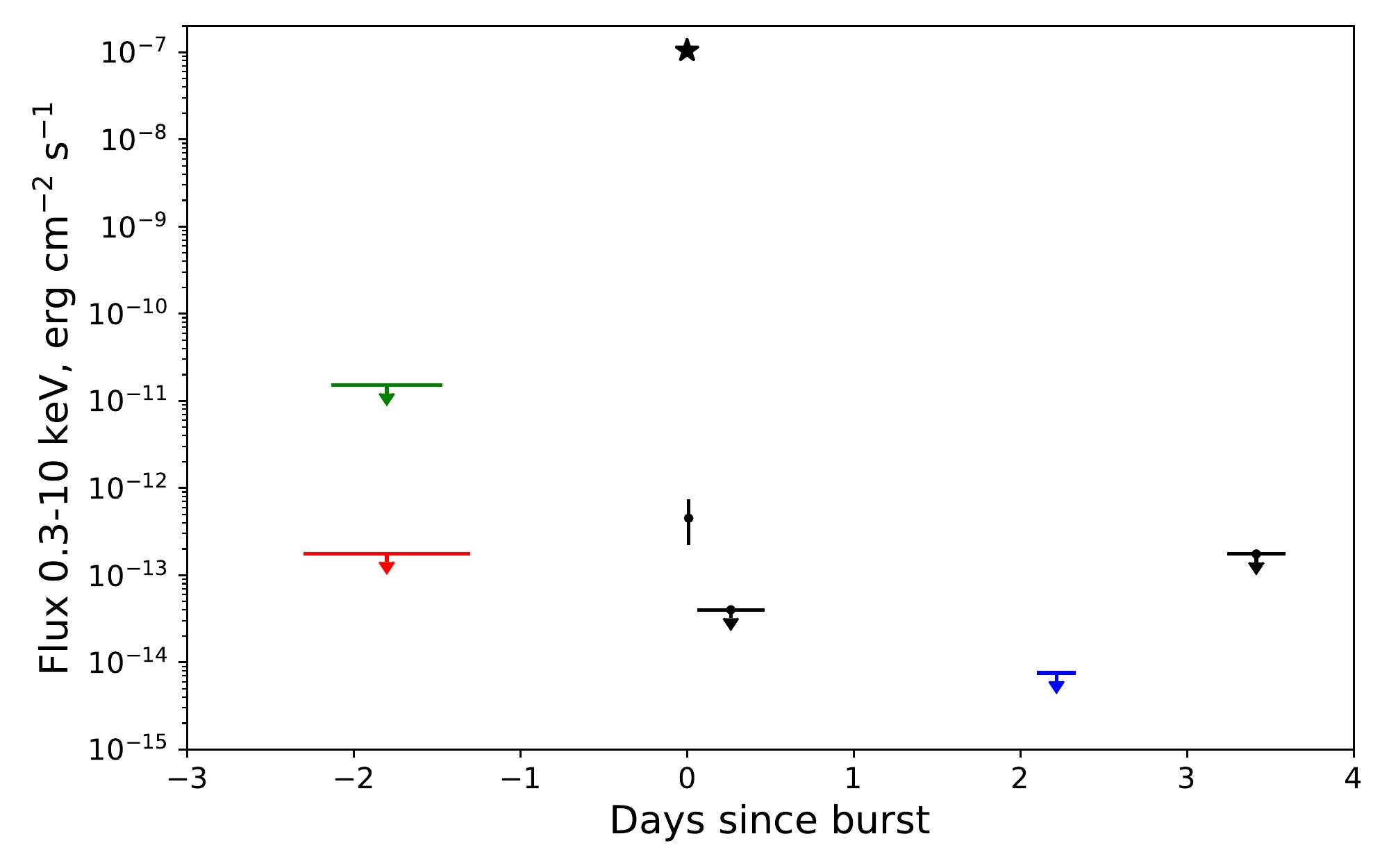}
     \caption{ Observed fluxes from GRB210919A, translated to the 0.3-10 keV energy band. The black star corresponds to the main pulse detected by \sbat, black points indicate the \sxrt\, afterglow detection and follow-up upper limits. Upper limits obtained before the burst are from \ero\, (red) and \art\, (green); the upper limit from the follow-up {\it Chandra} observation is shown in blue.}
     \label{fig:lc}
\end{figure*}

Overall limits on pre-merger X-ray emission derived from \srg\, telescopes are shown in Fig.~\ref{fig:spe}, and the combined lightcurve is presented in Fig.~\ref{fig:lc}.

\section{Upper limits on luminosity of pre-merger BNS}

As it was stated earlier, there is no direct measurement of the redshift of GRB210919A. The quality of the high-energy X-ray ($\gtrsim$100 keV) data is also insufficient to use Amati relation or similar \citep[see ][and references]{minaev20} to assess the distance. Assuming that the discovered galaxy group at $z=0.2411$ ($D \simeq 1.2$ Gpc)   \citep{2021GCN.30934....1O, LBTz22} is, indeed, hosts the GRB progenitor system, we can place upper limits on its isotropic X-ray luminosity: $L_{\ero\,}\le 10^{43}$ erg s$^{-1}$, $L_{\art\,}\le  10^{45}$ erg s$^{-1}$.

Now we can estimate the total isotropic-equivalent energy of the GRB as $E_{iso}\approx 10^{50}$ erg, using measured fluence and extrapolating \sbat\, spectrum. From proposed relation between $E_{iso}$ and a viewing angle \citep{salafia19} one can assume that the initial binary system was seen nearly edge-on. For the standard 1.4 $M_\odot$ masses of the components, the two-day time before the coalescence corresponds to an orbital separation of $a_0\sim 100 R_{NS}\sim 10^8$ cm. This separation can be smaller than the light cylinder of one of the components, $R_l=c/\omega=5\times 10^9(P/1\mathrm{s})$~cm ($P$ is the NS spin period). In this case, the expected electromagnetic power is $L_\mathrm{em}\sim 10^{38}(B_s/10^{15}\mathrm{G})^2(a_0/10^8\mathrm{cm})^{-7}$ erg s$^{-1}$ \citep{hansen01}, where $B_s$ is the NS surface magnetic field. If the NS spin period is shorter, there may be the case where the NS magnetosphere size is smaller than the orbital separation. This configuration was considered in \cite{wang18}. In the most favourable case of anti-aligned magnetic dipole moments of two NSs, the expected electromagnetic power is $L_\mathrm{em}\sim 4\times 10^{41}(B_s/10^{12}\mathrm{G})^2(a/10^8\mathrm{cm})^{-2}$ erg s$^{-1}$. While the total power in this case can be commensurable with the \ero\, upper limits, the expected spectra are too soft to produce X-ray emission. Thus, the obtained X-ray upper limits are too loose to constrain the possible physical properties of the putative binary NS system two days before the merging.
 
\section{Discussion}

Detection of pre-merger emission from merging BNS is a tempting, although complex observational task. However, thanks to its observational strategy telescopes on board \srg\, could detect such events during the all-sky survey.   
We have analysed observation of sGRB GRB210919A, that was observed by {\it SRG} less than two days before the merger. We have obtained, for the first time, upper limits on pre-merger X-ray emission on day-length timescales: assuming that the merger happened in galaxy group at $z=0.2411$ upper limits are $L_{\ero\,}\le 10^{43}$ erg s$^{-1}$ and $L_{\art\,}\le  10^{45}$ erg s$^{-1}$. 

We have estimated, that during the 4 year survey \srg\, could observe about 0.2 sGRB serendipitously less than a day before the merger.

\section*{ACKNOWLEDGEMENTS}

This work is based on the data from Mikhail Pavlinsky ART-XC and eROSITA X-ray instruments on board the SRG observatory. The SRG observatory was built by Roskosmos in the interests of the Russian Academy of Sciences represented by its Space Research Institute (IKI) in the framework of the Russian Federal Space Program, with the participation of the Deutsches Zentrum für Luft- und Raumfahrt (DLR). The ART-XC team thanks the Russian Space Agency, Russian Academy of Sciences and State Corporation Rosatom for the support of the \srg\ project and \art\ telescope and the Lavochkin Association (NPOL) with partners for the creation and operation of the \srg\ spacecraft (Navigator). The eROSITA X-ray telescope was built by a consortium of German Institutes led by MPE, and supported by DLR.  The science data are downlinked via the Deep Space Network Antennae in Bear Lakes, Ussurijsk, and Baykonur, funded by Roskosmos. The eROSITA data used in this work were processed using the eSASS software system developed by the German eROSITA consortium. 

Authors are grateful to referees for critical remarks.
This work was supported by the RFBR grant 19-29-11029. 
Work of KAP (interpretation of the results) was supported by Kazan Federal university program "Priority-2030".   

\bibliographystyle{mnras} 
\bibliography{reflist_eng} 

\label{lastpage}
\end{document}